\documentclass[aps,prd,twocolumn,superscriptaddress,nofootinbib]{revtex4-1}

\usepackage{latexsym}
\usepackage{amsmath}
\usepackage{amssymb}
\usepackage{amsfonts}
\usepackage{bm}
\usepackage{physics}
\usepackage{mathrsfs}

\usepackage{color}
\definecolor{purple}{rgb}{0.5,0,0.5}
\definecolor{blue}{rgb}{0.0,0,0.9}
\definecolor{prdblue}{rgb}{0.133,0.118,0.498}
\usepackage[colorlinks=true, pdfstartview=FitV, linkcolor=prdblue, citecolor= prdblue, urlcolor=prdblue]{hyperref}

\usepackage{supertabular} 
\usepackage{placeins}
\usepackage{epsfig}
\usepackage{graphicx}

\begin{document}


\title{$D^* D \pi$ and $B^* B \pi$ couplings from Dyson-Schwinger equations framework}


\author{Y.-Z. Xu}
\email[]{yinzhen.xu@dci.uhu.es}
\affiliation{Departmento de Ciencias Integradas, Universidad de Huelva, E-21071 Huelva, Spain.}
\affiliation{Departamento de Sistemas F\'isicos, Qu\'imicos y Naturales, Universidad Pablo de Olavide, E-41013 Sevilla, Spain}

\author{K. Raya}
\email[]{khepani.raya@dci.uhu.es}
\affiliation{Departmento de Ciencias Integradas, Universidad de Huelva, E-21071 Huelva, Spain.}


\date{\today}

\begin{abstract}

Employing a unified Dyson-Schwinger/Bethe-Salpeter equations approach, we calculate the strong decay couplings $D^* D \pi$ and $B^* B \pi$ within the so-called impulse-approximation in the moving frame. The $B^* B \pi$ estimation is reported for the first time based on a Poincaré invariant computation of the associated Bethe-Salpeter amplitudes. Our predictions yield $g_{D^* D \pi}=16.22_{-0.01}^{+0.03}$ and $g_{B^* B \pi}=40.09_{-1.37}^{+1.51}$, along with corresponding static strong couplings of $\hat{g}_D=0.55_{-<0.01}^{+<0.01}$, $\hat{g}_B=0.50_{-0.02}^{+0.02}$, which are consistent with recent experimental and lattice data.

\end{abstract}


\maketitle


\section{INTRODUCTION}
\label{sec:intro}

The study of the static and dynamical properties of the bound-states emerging from the strong interactions described by quantum chromodynamics (QCD), namely hadrons,  is undoubtedly a fundamental topic in modern physics. Among these properties, the strong couplings of heavy-light pseudo-scalar and vector mesons with the pion turn out to be essential hadronic parameters for heavy flavor physics\,\cite{HFLAV:2022esi}. The $D^* D \pi$ coupling ($g_{D^*D\pi}$) can be extracted from the total width of the $D^*$ meson by combining the branching fractions of the $D^*\rightarrow D\pi$ decays. On the contrary, the analogous $B^* B \pi$ case cannot be directly measured due to an insufficiency of phase-space for the $B^*\rightarrow B\pi$ decay to occur\,\cite{Flynn:2015xna}. Its determination still remains important whatsoever\,\cite{Belyaev:1994zk,Khodjamirian:2020mlb}. For instance, because it approaches more closely to the heavy-quark limit than $D^*D\pi$, enabling a more precise identification of the so called static strong coupling $\hat{g}$ that characterizes the interaction of heavy-light mesons with the pion. The latter is vital within the context of heavy meson chiral perturbation theory\,\cite{Burdman:1992gh,Wise:1992hn,Yan:1992gz,Flynn:2015xna}.\par  
In the past few decades, the determination of the $g_{H^*H\pi}$ couplings has attracted growing attention and various methods have been applied, for example, lattice QCD (lQCD)\,\cite{Flynn:2015xna,Bernardoni:2014kla,Can:2012tx,Becirevic:2012pf,Detmold:2012ge,Detmold:2011bp}, heavy quark effective theory (HQET)\,\cite{Cheung:2014cka}, QCD sum rules (SR)\,\cite{Belyaev:1994zk,Khodjamirian:2020mlb, Duraes:2004uc, Navarra:2001ju} and the combined Dyson-Schwinger and Bethe-Salpeter equations framework (DSEs/BSEs)\,\cite{Jarecke:2002xd,Mader:2011zf,El-Bennich:2010uqs,daSilveira:2022pte}. From the diverse set of approaches, the latter stands out for its capability to provide a non-perturbative and Poincaré-covariant framework, capable of simultaneously capturing essential traits of QCD, such as confinement and dynamical chiral symmetry breaking (DCSB). Consequently, it has been exploited for over thirty years to scrutinize the different vacuum and in-medium facets of QCD as well as several hadron properties  (see e.g. Refs.\,\cite{Roberts:1994dr,Maris:1997hd,Maris:1997tm,Maris:1999nt,Jarecke:2002xd,El-Bennich:2010uqs,Mader:2011zf,Qin:2011dd,Qin:2011xq,Raya:2015gva,Eichmann:2016yit,Fischer:2018sdj,
Qin:2019oar,Chen:2019otg,Xu:2019ilh,Serna:2020txe,Xu:2020loz,Xu:2021lxa,Xu:2021mju,Serna:2022yfp,daSilveira:2022pte,Li:2023zag,Xu:2023vlt,Xu:2023izo,Xu:2024vkn,Raya:2024ejx,Xu:2024fun,Shi:2024laj,Gao:2024gdj,Xu:2025hjf,Xu:2025sxw}).\par 
The investigation of strong decays of light vector mesons, based upon a DSEs/BSEs treatment that yields Poincar\'e invariant Bethe-Salpeter amplitudes (BSAs), can be traced back to Ref.\,\cite{Jarecke:2002xd}, where the $\rho\to\pi\pi$, $\phi\to KK$ and $K^*\to K\pi$ decays were computed for the very first time within this approach. Results concerning the $\rho\pi\pi$ case were reanalyzed afterwards in Ref.\,\cite{Mader:2011zf}. Concurrent investigations that included $D^*$ and $B^*$ mesons were carried out in\,\cite{El-Bennich:2010uqs}. These would constitute the latest explorations on the $B^*$ meson case under the DSEs/BSEs framework. Nonetheless, Ref.\,\cite{El-Bennich:2010uqs} utilizes a one-covariant model for the corresponding BSAs, which, while illustrative, compromises Poincar\'e invariance by truncating the structure of such\,\cite{Maris:1997hd, Maris:1997tm}. An updated account of these results, that leaves out the $B^*$ mesons, is provided in\,\cite{daSilveira:2022pte}. Therein, a complete Dirac structure for the BSAs is considered, along with a truncation scheme similar in essence to the well-known rainbow-ladder (RL) approximation, which features a modified effective interaction to better capture the significant flavor asymmetry observed in heavy-light mesons\,\cite{Serna:2020txe,Serna:2022yfp,Chen:2019otg}. While the $g_{D^*D\pi}$ estimation obtained in \,\cite{daSilveira:2022pte} aligns well with experimental expectations\,\cite{ParticleDataGroup:2022pth}, such examination reveals some technical complexities in addressing heavy-light systems.\par   
Recently, significant progress has been made in studying the properties and dynamics of heavy-light mesons using Poincaré-invariant BSAs. This research encompasses various areas, including mass spectra, decay processes, distribution amplitudes, electromagnetic form factors, and generalized parton distributions\,\cite{Chen:2019otg,Qin:2019oar,Serna:2020txe,Serna:2022yfp,daSilveira:2022pte,Xu:2024fun,Shi:2024laj,Gao:2024gdj,Xu:2025hjf,Xu:2025sxw}. Building upon our previous work on the electromagnetic form factors of heavy-light mesons,\,\cite{Xu:2024fun}, in the present work we extend the investigation to include $D^* D \pi$ and $B^* B \pi$ couplings. In our analysis, the BSAs are solved directly based upon a moving frame formulation of the corresponding BSEs,\,\cite{Bhagwat:2006pu}, thus eliminating the fitting/extrapolation errors of the BSAs that appeared in previous similar studies in the evaluation of the triangle diagram. Notably, a prediction for the $B^*B\pi$ coupling is reported for the first time based on the Poincaré invariant determination of the BSAs. Our results are consistent with recent experimental extractions and available lattice data. \par 
This manuscript is organized as follows: In Sec.\,\ref{sec:2}, we introduce the essential components for the computation of strong decays within the DSEs/BSEs framework. The obtained numerical results for the $D^*D\pi$ and $B^*B\pi$ couplings are presented in Sec.\,\ref{sec:3}. We also examine how contributions from the various Dirac structures characterizing the BSAs impact the overall results. Finally, a brief summary is provided in Sec.\,\ref{sec:4}.


\section{The strong coupling within DSEs/BSEs framework}
\label{sec:2}

\subsection{The strong decays and the impulse approximation}
\begin{figure}
\centering 
\includegraphics[width=.35\textwidth]{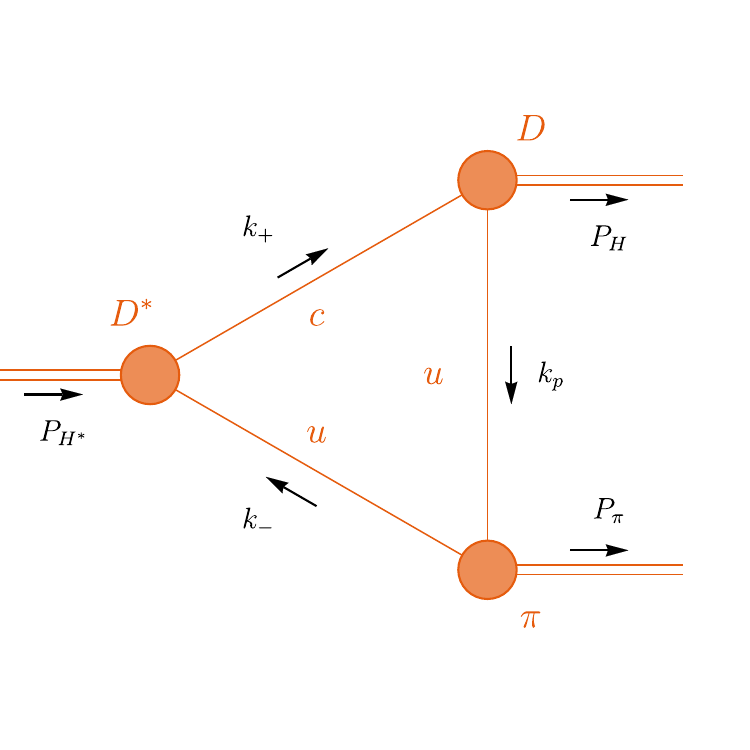}
\includegraphics[width=.35\textwidth]{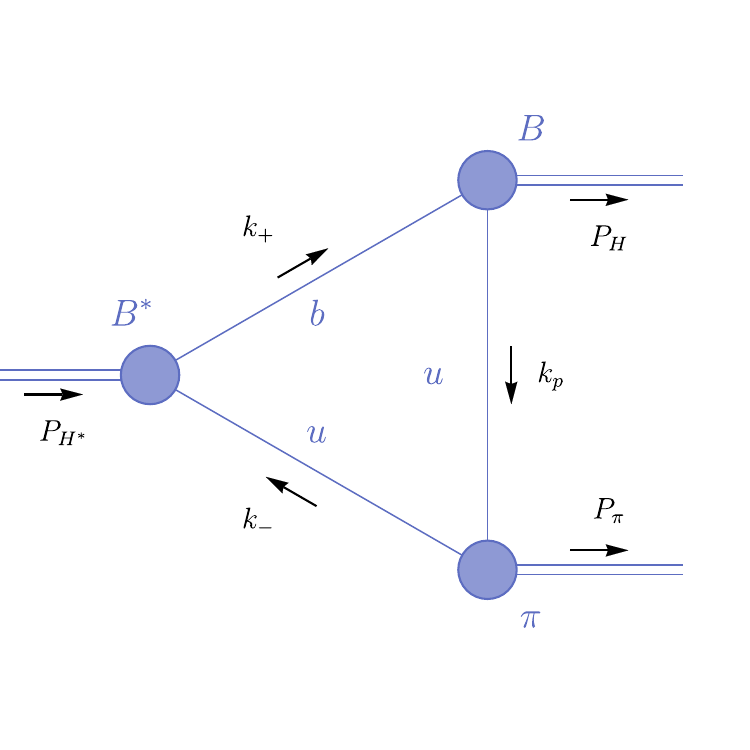}
\caption{\label{fig:sd} Upper panel: $D^*\rightarrow D\pi$; lower panel: $B^*\rightarrow B\pi$. The solid lines represent the dressed-quark propagator, the shaded circles denote the meson's BSAs, and the black arrows indicate the direction of momentum. We work in the $\mathrm{SU}(2)$ isospin limit, hence $u = d$ quark.}
\end{figure}

The strong $H^* H \pi$ coupling, with $H=D,B$, can be defined by the on-shell matrix element \cite{Belyaev:1994zk,El-Bennich:2010uqs}
\begin{align}
  \left\langle H\left(P_{H}\right) \pi(P_{\pi}) \mid H^*\left(P_{H^*}, \lambda\right)\right\rangle=g_{H^* H \pi} \epsilon^\lambda \cdot P_{\pi},
\end{align}
where $\epsilon_\mu^{\lambda}$ is the polarization vector of the $H^*$ meson, and $P_{\pi}$, $P_{H}$, $P_{H^*}$ are the four-momenta of the pion, pseudo-scalar meson, and vector meson, respectively. We work in Euclidean space and these four-momenta satisfy the following on-shell conditions:
\begin{align}
  P_{\pi}^2=-M^2_\pi;\  P_{H}^2=-M^2_H;\  P_{H^*}^2=-M^2_{H^*}\,.
\end{align}
Within the widely used impulse approximation, the strong decay processes under examination can be evaluated from a triangle diagram that turns out to be fully characterized by the dressed quark propagators ($S$) and bound-state BSAs ($\Gamma_H$) as follows:
\begin{align}
 \tilde{\Lambda}_{\mu} =\, & N_c \text{Tr}\int^{\Lambda}\frac{d^4 \tilde{k}}{(2\pi)^4}  S^{\bar{f}}\left(k_p\right)\Gamma_{H}^{(\text{out})}\left(k_{p}, k_{+}\right) S^g\left(k_{+}\right) \nonumber \\
 & \Gamma_{\mu,H^*}^{\text{(in)}}\left(k_{+}, k_{-}\right) S^g\left(k_{-}\right) \Gamma_{\pi}^{(\text {out})}\left(k_{-}, k_p\right) ,
\label{eq.sd}
\end{align}
with $ \epsilon^\lambda\cdot \tilde{\Lambda} = g_{H^*H\pi} \epsilon^\lambda \cdot P_{\pi}$. Here $\text{Tr}$ stands for a trace over Dirac indices, $N_c=3$, and $\Lambda$ represents a regularization scale. In Fig.\,\ref{fig:sd}, a more intuitive pictorial representation of the above mathematical expression is shown. The computation of the components entering Eq.\,\eqref{eq.sd} within the present DSEs/BSEs framework shall be detailed afterwards.
\par 
What remains now is to define the kinematics to be adopted in Eq.\,\eqref{eq.sd}. For the momenta of quarks and mesons, we further define 
\begin{subequations}
\begin{align}
P_{\pi}&=-\tilde{P}+(1-\alpha)\tilde{Q},\ P_{H}=\tilde{P}+\alpha \tilde{Q},\ P_{H^*}=\tilde{Q};\\
k_{+}&=\tilde{k}+\alpha\tilde{Q},\ k_{-}=\tilde{k}-(1-\alpha)\tilde{Q},\ k_p=\tilde{k}-\tilde{P},
\end{align}
\label{eq.krkskp}
\end{subequations}
where $\alpha$ is a momentum partitioning parameter associated to the $H^*$ meson, $\tilde{k}$ represents the integral momentum, and 
\begin{subequations}
\begin{align}
  \tilde{Q} = (0,0,0,1)\tilde{Q}_s,\ \tilde{P} = (0,0,\sqrt{1-z^2},z)\tilde{P}_s,
\end{align}
with
\begin{align}
  \tilde{Q}_s & = iM_{H^*},\\
  \tilde{P}_s & = i\sqrt{M_{H}^2-\alpha \left((1-\alpha)
   M_{H^*}^2-M_{\pi}^2+M_{H}^2\right)},\\
   z&=\frac{(1-2 \alpha) M_{H^*}^2-M_{\pi}^2+M_{H}^2}{2 M_{H^*}
   \sqrt{M_{H}^2-\alpha \left((1-\alpha)
   M_{H^*}^2-M_{\pi}^2+M_{H}^2\right)}}\,.
\end{align}
\label{eq.frame}
\end{subequations}\par 
In our notation/coordinate convention, i.e., Eqs.\,(\ref{eq.krkskp})-(\ref{eq.frame}), $\tilde{Q}$ is the total momentum of the system \begin{align}
   \tilde{Q}=P_{H^*}=P_{H}+P_{\pi}\,.
\end{align}
Furthermore, if we consider the \(\rho \rightarrow \pi\pi\) process and set \(\alpha = 1/2\), \(\tilde{P}\) reduces to the relative momentum between the outgoing pions:  
\begin{align}
\left. \tilde{P} \right|_{\rho\pi\pi,\ \alpha=1/2}=(0,0,\kappa,0),\ \kappa^2=\frac{M_{\rho}^2}{4}-M_{\pi}^2\,,
\end{align}
which is consistent with the formulation employed in previous similar studies of \(g_{\rho\pi\pi}\) \cite{Mader:2011zf, Jarecke:2002xd}.\par 
We now proceed to discuss the evaluation of Eq.\,\eqref{eq.sd}. In this work, we employ the brute-force/moving-frame approach \cite{Bhagwat:2006pu,Xu:2019ilh,Xu:2023izo,Xu:2024fun,Xu:2024vkn}, in which the BSEs are solved in the frame where the external momenta of the quark legs are fixed by the triangle diagram, and the resulting BSAs, after normalization, are directly used as input in Eq.\,\eqref{eq.sd}. A complete introduction to this method is provided in Ref.\,\cite{Bhagwat:2006pu}. \par 
Therefore, tuning $\alpha$ in Eq.\,\eqref{eq.sd} induces corresponding changes to the external momenta of the quark legs in the associated BSEs (see Eq.\,\eqref{eq:hBSE} for details). This procedure allows one to circumvent the singularities of the quark propagator in the complex momentum plane. This treatment is particularly important for $D^*D\pi$ and $B^*B\pi$, where $\alpha$ is constrained to a narrow finite range that excludes 1/2. Accordingly, \(\alpha \neq 1/2\) is used in the entire calculation of the \(D^*D\pi\) and \(B^*B\pi\) processes in this work. A detailed discussion will be presented in the next subsection.\par 
In principle, due to Poincar\'e invariance, observable quantities should not depend on \(\alpha\), as long as it is kept consistent throughout the entire calculation. As we will demonstrate, within the adjustable range of \(\alpha\), the value of \(g_{H^*H\pi}\) is indeed independent of the choice of \(\alpha\).
\subsection{Quark propagators from DSEs}
On general grounds, the dressed-quark propagators entering Eq.\,\eqref{eq.sd} are expressed in terms of two Dirac structures as  
\begin{equation}
  S^{-1}(k)= i\gamma \cdot p\,A(k^2) + B(k^2)\,,
\end{equation}
where $A(k^2)$ and $B(k^2)$ are scalar functions to be determined. The quark propagator satisfies the following gap equation:
\begin{align}
  S^{-1}(k)=&Z_2i \gamma \cdot k + Z_4 m \nonumber \\ 
  &+ Z_{1} \int^{\Lambda} \frac{d^4 q}{(2 \pi)^4}  g^2 D_{\mu \nu}(l) \frac{\lambda^a}{2} \gamma_\mu S(q) \frac{\lambda^a}{2} \Gamma_\nu(k,q)\,,
  \label{eq.DSE}
\end{align} 
where $l=k-q$, $m$ is current-quark mass, and $Z_{1,2,4}$ are the renormalization constants. In this work we employ a mass-independent momentum-subtraction renormalization scheme \footnote{The mass-independent momentum subtraction renormalization is a variant of the widely used momentum subtraction scheme. Detailed discussions and applications of the latter in DSEs are available in Refs.~\cite{Maris:1997tm,Blank:2011qk}. The key difference between the two schemes lies in whether renormalization is performed in the chiral limit to fix the renormalization constants \cite{Chang:2008ec,Bashir:2012fs,Liu:2019wzj}.}, setting the renormalization scale at $\zeta = 19$ GeV 
\footnote{The use of 19 GeV as the renormalization scale can be traced back to Refs.\,\cite{Maris:1997tm,Maris:1999nt}, where it was chosen as it is large enough to be in the perturbative domain, allowing the one-loop asymptotic behavior of the quark propagator to be used as a check. Since then, this value has been widely adopted in DSE-based studies \cite{Qin:2011xq,Serna:2018dwk,Eichmann:2019bqf}. In line with our previous work \cite{Xu:2021mju,Xu:2024fun} and earlier investigations of strong decays within the DSEs/BSEs framework \cite{Jarecke:2002xd}, we likewise adopt 19 GeV as the renormalization scale in this study.}. In conjunction with dressed quark-gluon vertex $\Gamma_\nu(k,q) \rightarrow Z_2\gamma_\nu$ \cite{Binosi:2014aea}, which characterizes the rainbow-ladder (RL) truncation, we will adopt the following form for the gluon propagator:
\begin{equation}
  Z_1 g^2 D_{\mu \nu}(l)\Gamma_\nu(k,q)=Z_2^2 \mathcal{G}\left(l^2\right) \mathcal{P}_{\mu \nu}^T(l)\gamma_\nu=\mathcal{D}_{\text{eff}}\gamma_\nu\,,
  \label{eq.gluon}
\end{equation}
where $\mathcal{P}_{\mu \nu}^T(l)=\delta_{\mu \nu}-{l_\mu l_\nu}/l^2$ is the standard transverse projection operator, and an Abelian Ward identity is enforced that leads to $Z_1=Z_2$, which at one loop corresponds to neglecting the contribution of the three-gluon vertex to $\Gamma_\nu(k, q)$ \cite{Maris:1997tm,Serna:2018dwk,daSilveira:2022pte}. For the effective interaction, we employ the well-known Qin-Chang model\,\cite{Qin:2011dd,Qin:2011xq}:
\begin{subequations}
  \label{eq.qc}
\begin{align}
\frac{\mathcal{G}(l^2)}{l^2}&=\mathcal{G}^{\text{IR}}(l^2)+\frac{8\pi^2\gamma_{m}\mathcal{F}(l^2)}{\ln[\tau+(1+l^2/\Lambda^2_{\text{QCD}})^2]},\\ 
\mathcal{G}^{\text{IR}}(l^2)&=D\frac{8\pi^2}{\omega^4}e^{-l^2/\omega^2}\,.
  \end{align}
\end{subequations}
The pieces entering the ultraviolet-dominating contribution are defined as follows: $\mathcal{F}(l^2)=\{1-\exp[(-l^2/(4m_t^2)]\}/l^2$, $m_t=0.5$\,GeV, $\tau=e^2-1$, $\Lambda_{\text{QCD}}=0.234$\,GeV, $\gamma_{m}=12/25$\,\cite{Xu:2021mju}. For the infrared component, in line with Refs.\,\cite{Xu:2024fun,Shi:2024laj} we choose $(D\omega)_{u/d}=(0.82\ \text{GeV})^3$, $(D\omega)_c=(0.66\ \text{GeV})^3$, $(D\omega)_b=(0.48\ \text{GeV})^3$, with $\omega_{u/d}=0.5$ GeV, $\omega_{c,b}=0.8$ GeV. More details concerning Eqs.\,(\ref{eq.DSE}-\ref{eq.qc}) are found through Refs.\,\cite{Roberts:1994dr,Maris:2003vk,Maris:1997tm,Qin:2011dd,Maris:1999nt}.\par
\begin{figure*}
\centering 
\includegraphics[width=.3\textwidth]{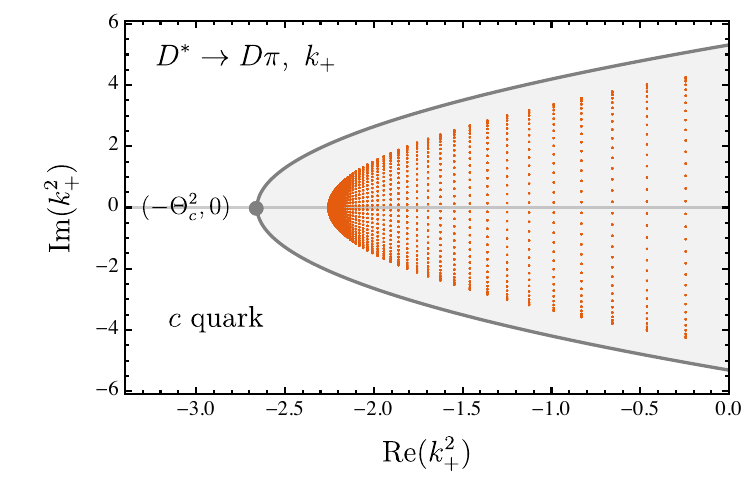}
\includegraphics[width=.3\textwidth]{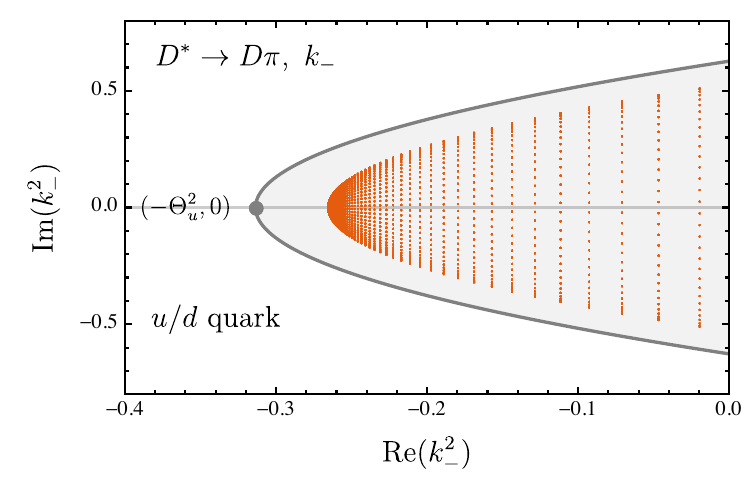}
\includegraphics[width=.3\textwidth]{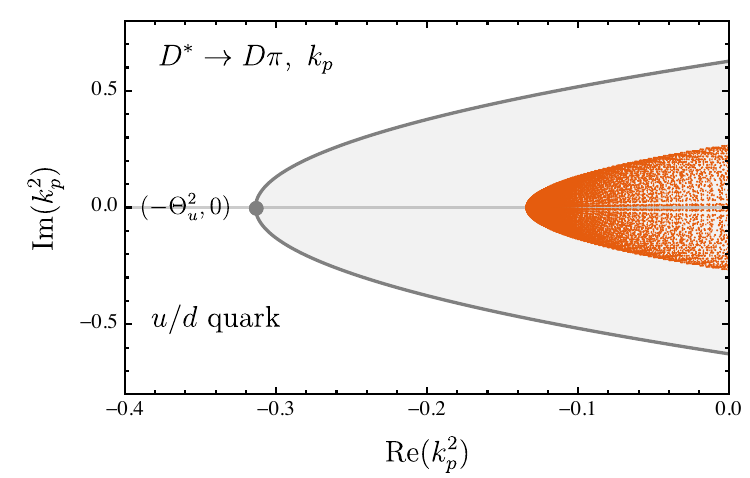}
\includegraphics[width=.3\textwidth]{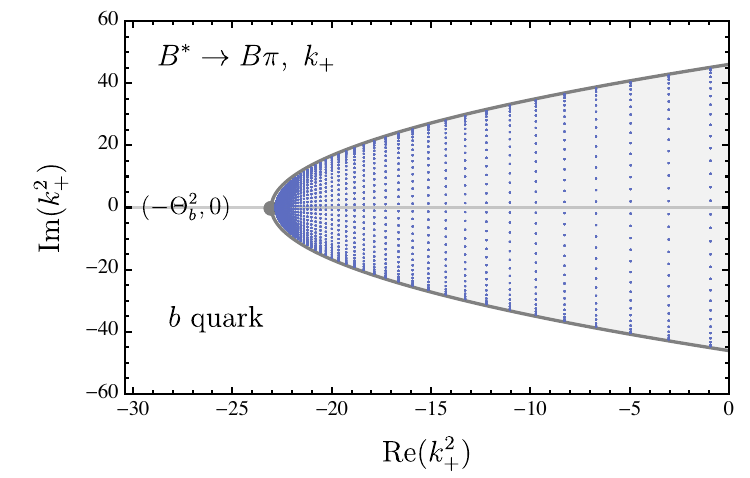}
\includegraphics[width=.3\textwidth]{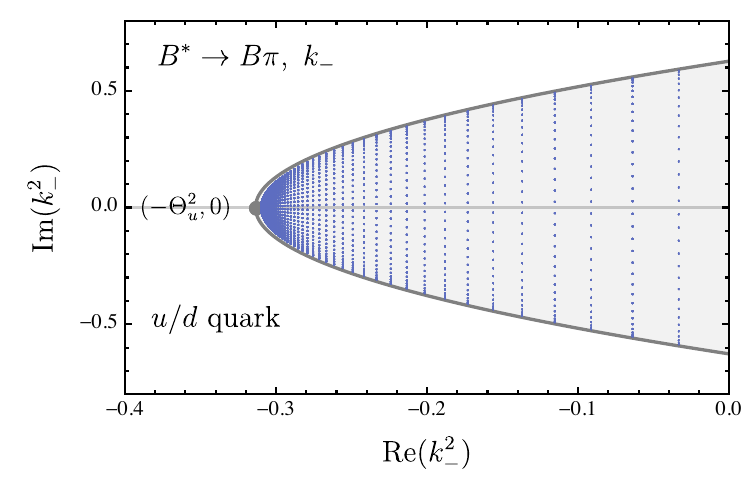}
\includegraphics[width=.3\textwidth]{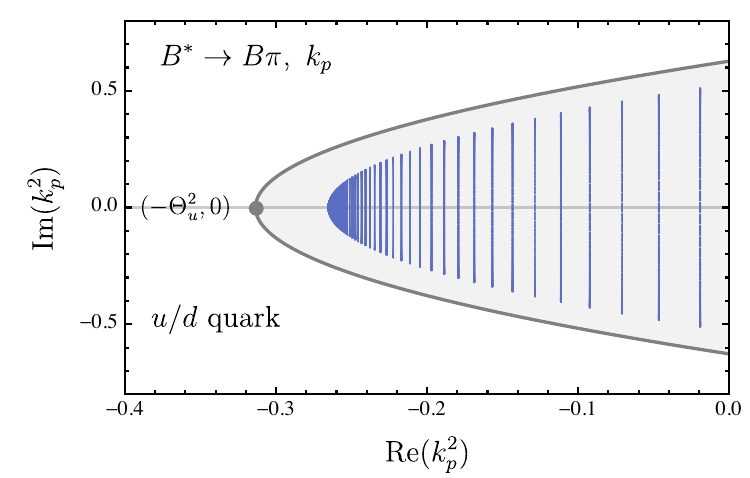}
\caption{\label{fig:sd_quark} Required evaluation domain for the quark propagator. The colored points indicate the actual required values of $k_{-}^2$, $k_{+}^2$, and $k_p^2$ in Eq.\,\eqref{eq.sd} under the optimal $\alpha$; only those with negative real parts are shown. The gray area represents the region that is computable.}
\end{figure*}

 
It is worth stressing that, in solving the bound-state problem, Eq.\,\eqref{eq.DSE} usually needs to be evaluated in the complex plane to ensure the relative momentum $l$ in the interaction model is real,\,\cite{Fischer:2005en,Rojas:2014aka}. Notably, $k^2$ is constrained by a parabola (see Fig.\,\ref{fig:sd_quark}, gray region)
\begin{equation}
	\text{Re}(k^2) \geq -\Theta^2 +\frac{\text{Im}^2(k^2)}{4 \Theta^2},
	\label{eq.parabola}
\end{equation}
where $(- {\Theta}^2, 0)$ is the vertex of parabola. Once the quark propagator on this region is determined, its values at any interior point of the contour can be directly obtained using the Cauchy integral theorem \cite{Sanchis-Alepuz:2017jjd}. However, poles in the complex plane compromise the applicability of this technique: they cannot be enclosed by the integration contour. For quarks of different flavors, the positions of the singularities vary\,\cite{Windisch:2016iud}, so the maximum value adopted by the vertex also differs. In this work, we obtain $\Theta_u=0.56$ GeV, $\Theta_c=1.63$ GeV, $\Theta_b=4.8$ GeV. \par 
With the above in mind, note that the integral momentum $\tilde{k}$ in Eq.\,\eqref{eq.sd} can be conveniently written as
\begin{equation}
	\tilde{k}=(0,\sin \iota \sin \theta,\cos \iota \sin \theta, \cos \theta )|\tilde{k}|\,,
\end{equation}
and, accordingly,
\begin{subequations}
	\begin{align}
	\text{Re}(k^2_+) &= -\alpha^2 M_{H^*}^2 +\frac{\text{Im}^2(k^2_+) \sec^2 \theta}{4 M_{H^*}^2 \alpha^2},\   \theta \in [0, \pi]\,,\\
	\text{Re}(k^2_-) &= -(1-\alpha)^2 M_{H^*}^2 +\frac{\text{Im}^2(k^2_-) \sec^2 \theta}{4 M_{H^*}^2 (1-\alpha)^2},\   \theta \in [0, \pi]\,. 	
\end{align}
\label{eq.krks}
\end{subequations}
The resemblance between Eq.\,\eqref{eq.parabola} and Eq.\,\eqref{eq.krks} is readily noticeable; in the latter case, this corresponds to making the association with the rest frame of the $H^*$ meson. Although the physical observables do not depend on $\alpha$, in the actual calculation, the selection of $\alpha$ should ensure that $k_{+}^2$ and $k_{-}^2$ remain within a computable region  
\begin{align}
   \alpha^2 M_{H^*}^2 < \Theta^2,\ (1-\alpha)^2 M_{H^*}^2 < \Theta^2, 
\end{align}
such that
\begin{align}
  1-\Theta_{\bar{g}}/M_{H^*}<\alpha<\Theta_f/M_{H^*}\,.
  \label{eq.alpha}
\end{align}
Consequently, the optimal value of $\alpha $ can be defined as $\Theta_f / (\Theta_f + \Theta_{\bar{g}})$, which corresponds to the maximum computable mass $\Theta_f + \Theta_{\bar{g}}$ in the calculation of meson mass spectra \cite{Rojas:2014aka}. In this work, the obtained optimal \(\alpha\) are 0.744 and 0.896 for the \(D^*\) and \(B^*\) mesons, respectively. The central value of our prediction for \( g_{H^*H\pi} \) is determined at this optimal \(\alpha\), while its uncertainty arises from variations of \(\alpha\) within the range specified in Eq.\,(\ref{eq.alpha}). \par 
As for $k_p^2$ concerns, under the current notation, it cannot be expressed in a simple form similar to Eq.\,\eqref{eq.krks}. Nevertheless, numerical results show that $k_p^2$ remains within the computable region in the calculations of $D^*\rightarrow D\pi$ and $B^*\rightarrow B\pi$. Fig.\,\ref{fig:sd_quark} displays the specific values required for $k_{-}^2$, $k_{+}^2$, and $k_p^2$ when their real parts are negative, given the optimal choice of $\alpha$.
\subsection{Meson's BSAs from BSEs}
The next components entering Eq.\,\eqref{eq.sd} are the BSAs of the corresponding mesons. In the brute-force/moving-frame approach, one needs to solve the BSEs as shown below \cite{Bhagwat:2006pu}:
\begin{subequations}
\begin{align}
  \Gamma^{f\bar{g}}_{\mu,H^*}\left(k_{+}, k_{-}\right)=&\int^{\Lambda} \frac{d^4 q}{(2 \pi)^4} K^{f \bar{g}}(q, \tilde{k} ; P_{H^*}) S^f\left(q_{+}\right)\nonumber
  \\ & \Gamma_{\mu,H^*}^{f \bar{g}}\left(q_{+}, q_{-}\right) S^g\left(q_{-}\right);\\
   \Gamma^{f\bar{g}}_H\left(k_{p}, k_{+}\right)= & \int^{\Lambda} \frac{d^4 q}{(2 \pi)^4} K^{f \bar{g}}(q, \tilde{k} ; P_{H}) S^f\left(q_{p}\right) \nonumber \\ &\Gamma^{f \bar{g}}_{H}\left(q_{p}, q_{+}\right) S^g\left(q_{+}\right);\\
   \Gamma^{f\bar{g}}_{\pi}\left(k_{-}, k_{p}\right) =& \int^{\Lambda} \frac{d^4 q}{(2 \pi)^4} K^{f \bar{g}}(q, \tilde{k} ; P_{\pi}) S^f\left(q_{-}\right)\nonumber \\  & \Gamma^{f \bar{g}}_{\pi}\left(q_{-}, q_{p}\right) S^g\left(q_{p}\right),
\end{align}
\label{eq:hBSE}
\end{subequations}

with $f$ and $g$ labeling the quark and antiquark flavors, and $K^{f\bar{g}}$ being the two-body interaction kernel. This shall be discussed afterwards. In Eq.\,\eqref{eq:hBSE}, the external momenta of the quark legs, \(k_+\), \(k_-\), and \(k_p\), are exactly the same as those in Eq.\,\eqref{eq.sd}, and \(q_{+} = q + \alpha \tilde{Q},\ q_{-} = q - (1 - \alpha) \tilde{Q},\ q_p = q - \tilde{P}\), where \(q\) is the integration momentum \footnote{Since the same $\alpha$ (with $\alpha \ne 1/2$) is consistently employed throughout this work, the relative momenta between the quark and the antiquark, namely $(k_+ + k_-)/2$, $(k_p + k_+)/2$, and $(k_- + k_p)/2$, are complex in both the triangle diagram and the BSE. In contrast, the momenta in the gluon model within the BSE kernel, i.e., $k_+ - q_+$, $k_p - q_p$, and $k_- - q_-$, all reduce to $\tilde{k} - q$ and are therefore real \cite{Bhagwat:2006pu}.}.  Accordingly, after normalization, the resulting \(\Gamma_{\mu,H^*}(k_{+}, k_{-})\), \(\Gamma_H(k_{p}, k_{+})\), and \(\Gamma_{\pi}(k_{-}, k_{p})\) are directly used as input in Eq.\,\eqref{eq.sd}. Further details on this point, such as the alignment of the integration grid, are presented in the appendix of Ref.\,\cite{Bhagwat:2006pu}. \par  
Having specified the kinematics, let us now discuss the tensor structure of the BSAs. These have a general decomposition in terms of Dirac covariants as follows:
\begin{equation}
\Gamma\left(k_{\rm{out}},k_{\rm{in}}\right)=\sum_{i=1}^N \tau_i(k, P) F_i (k, P)\,,
    \label{eq:hBSEbasis}
\end{equation}
where \(k_{\rm{out}}\) and \(k_{\rm{in}}\) denote the external momenta of the quark legs, \(P\) is the total momentum, and \(k\) is related to the relative momentum of the quark \cite{Bhagwat:2006pu}. The \(\tau_i(k, P)\) are the basis tensors, and \(F_i(k, P)\) are the associated scalar functions. For the pseudo-scalar mesons we choose\,\cite{Qin:2011xq}:
\begin{subequations}
\label{eq:basis}
\begin{align}
\tau_{0^{-}}^1&=i \gamma_5,& &\tau_{0^{-}}^3=\gamma_5 \gamma \cdot k k \cdot P, \\
\tau_{0^{-}}^2&=\gamma_5 \gamma \cdot P,& &\tau_{0^{-}}^4=\gamma_5 \sigma_{\mu \nu} k_\mu P_\nu\,,
\end{align}
and, for their vector meson counterparts,
\begin{align}
& \tau_{1^{-}}^1=i \gamma_\mu^T  \\
& \tau_{1^{-}}^2=i\left[3 k_\mu^T \gamma \cdot k^T-\gamma_\mu^T k^T \cdot k^T\right] \\
& \tau_{1^{-}}^3=i k_\mu^T k \cdot P \gamma \cdot P \\
& \tau_{1^{-}}^4=i\left[\gamma_\mu^T \gamma \cdot P \gamma \cdot k^T+k_\mu^T \gamma \cdot P\right]\\
&\tau_{1^{-}}^5=k_\mu^T, \\
&\tau_{1^{-}}^6=k \cdot P\left[\gamma_\mu^T \gamma^T \cdot k-\gamma \cdot k^T \gamma_\mu^T\right], \\  
&\tau_{1^{-}}^7=\left(k^T\right)^2\left(\gamma_\mu^T \gamma \cdot P-\gamma \cdot P \gamma_\mu^T\right)-2 k_\mu^T \gamma \cdot k^T \gamma \cdot P, \\
&\tau_{1^{-}}^8=k_\mu^T \gamma \cdot k^T \gamma \cdot P.
\end{align}
\end{subequations}
with $V_\mu^T=V_\mu-P_\mu(V \cdot P) / P^2$. Note that if the complete structure of the BSAs is not retained, the Poincaré invariance, a feature of the direct application of DSEs/BSEs to the calculation of hadron properties, would be compromised\,\cite{El-Bennich:2010uqs,Maris:1997tm}. 
 \par  
The final piece completing the BSE, Eq.\,\eqref{eq:hBSE}, is the two-body interaction kernel $K^{f\bar{g}}$. In the standard RL approximation with a single-flavor gluon model, the interaction kernel is given by \cite{Xu:2021mju}:
\begin{equation}
\label{eq:kernel.RL}
  K^{f\bar{g}}(q,\tilde{k};P) = \tilde{\mathcal{D}}^{f\bar{g}}_{\text{eff}} \frac{\lambda^a}{2} \gamma_\mu \otimes \frac{\lambda^a}{2} \gamma_\nu,\  \tilde{\mathcal{D}}^{f\bar{g}}_{\text{eff}} = \mathcal{D}^f_{\text{eff}},
\end{equation}
where $ \mathcal{D}^f_{\text{eff}}$, defined in Eq.\,\eqref{eq.gluon}, describes the strength of interaction and decreases as the mass of quark increases because the quark dressing effects become suppressed \cite{Sultan:2018tet,Serna:2017nlr,Qin:2019oar}. In combination with the one-body kernel described in the previous section, Eq.\,\eqref{eq:kernel.RL} guarantees the axial-vector Ward-Takahashi identity (AV-WTI) is fulfilled, which ensures a massless pion in the chiral limit\,\cite{Munczek:1994zz}. From a numerical perspective, the degree of preservation can be further tested by assessing the commensurateness between the expected and obtained results for the generalized Gell-Mann-Oakes-Renner (GMOR) relation \cite{Rojas:2014aka,Chen:2019otg}:
\begin{equation}
   f_{0^-} M_{0^{-}} = \left(m_f+m_g\right) \rho_{0^{-}} \,.
   \label{eq.gmor}
\end{equation}
Clearly, in the above expression, $M_{0^-}$ corresponds to the pseudo-scalar meson's mass, while the quantity $\rho_{0^{-}}$ is defined as follows 
\begin{align}
  \rho_{0^{-}}= Z_4 N_c \text{Tr} \int^{\Lambda}\frac{d^4q}{(2\pi)^4} \gamma_5 S^f\left(q_{\rm{out}}\right) \Gamma_{0^{-}}^{f\bar{g}}(q_{\rm{out}},q_{\rm{in}}) S^{\bar{g}}\left(q_{\rm{in}}\right)\,;
\end{align}
and $f_{0^-}$ denotes the pseudo-scalar meson leptonic decay constant. Having properly normalized the BSAs\,\cite{Qin:2011xq}, the decay constants related to the pseudo-scalar and the vector can be obtained from the following expressions:
\begin{subequations}
	\begin{align}
  f_{0-}P_\mu = &Z_2 N_c \text{Tr} \int^{\Lambda} \frac{d^4q}{(2\pi)^4}\gamma_5 \gamma_\mu S^f\left(q_{\rm{out}}\right) \nonumber \\ 
  &\Gamma_{0^{-}}^{f g}(q_{\rm{out}},q_{\rm{in}}) S^g\left(q_{\rm{in}}\right),\\
  f_{1^{-}} M_{1^{-}} = & Z_2 N_c \text{Tr} \int^{\Lambda} \frac{d^4q}{(2\pi)^4} \gamma_\mu S^f\left(q_{\rm{out}}\right) \nonumber \\ 
  &\Gamma_{1^{-}}^{\mu, f g}(q_{\rm{out}}, q_{\rm{in}}) S^g\left(q_{\rm{in}}\right)\,.
\end{align}
\end{subequations}
\par 
\begin{table*}[!t]
\caption{\label{tab:meson} Produced masses and leptonic decay constants. It is worth noting that, in the coordinate convention of this work, the calculation of BSEs for vector mesons differs from the commonly used rest frame approach only in numerical details, and the results still correspond to the rest frame. Current quark masses are set as $m_{u,c,b}^{\xi=19\ \text{GeV}}=0.0033\ \text{GeV}, 0.854\ \text{GeV}, 3.682\ \text{GeV}$ \cite{Xu:2024fun} and the script $^\dagger$ denotes the fitting values from weight factor [see Eq.\,\eqref{eq.wRL}]. Here we work in the isospin-symmetric limit $m_u=m_d$. For comparison, we collect both experimental values (Expt.) and lQCD's results \cite{ParticleDataGroup:2022pth,Dowdall:2012ab,Cichy:2016bci,Lubicz:2017asp,Bazavov:2017lyh}.}
\begin{ruledtabular}
\begin{tabular}{l|llll|llll}
 Meson&\multicolumn{4}{c|}{Mass [GeV]} &\multicolumn{4}{c}{Leptonic decay constant [GeV]}\\
  & Expt. & lQCD & Rest frame & Moving frame &Expt. & lQCD & Rest frame& Moving frame\\ \hline
 $\pi$ & 0.138(1) & - & 0.135 & 0.135 & 0.092(1) & 0.093(1) & 0.095 & 0.095\\
 $D$&1.868(1)&1.868(3)& 1.868$^\dagger$ & 1.868 & 0.144(4)&0.150(4)& 0.140 & 0.140  \\
 $D^*$&2.009(1)&2.013(14)& 2.017 & ...  &-& 0.158(6)&0.160 & ... \\
 $B$&5.279(1)&5.283(8)&5.279$^\dagger$ &5.279 &0.133(18)&0.134(1)&0.123 &0.123 \\
 $B^*$&5.325(1)&5.321(8)&5.334 &... &-&0.131(5)&0.126 & ...
\end{tabular}
\end{ruledtabular}
\end{table*}

Over the past thirty years, Eq.\,\eqref{eq:kernel.RL} has been widely used in flavor symmetric/slightly asymmetric systems such as $u\bar{d}$, $u\bar{s}$, $c\bar{c}$ (e.g. \cite{Qin:2011dd,Ding:2015rkn}). However, for heavy-light systems, it is difficult to be applied because of the insufficiency of Eq.\,\eqref{eq:kernel.RL} to properly capture the marked flavor-asymmetry. A proven and rapidly developing approach involves using modified RL schemes that adjust the BSE's effective interaction to account for flavor asymmetry. This has allowed investigation of several aspects of heavy-light systems, ranging from mass spectra and decay constants, to electromagnetic form factors and many other quantities defined on the light-front\,\cite{Chen:2019otg,Qin:2019oar,Serna:2020txe,Serna:2022yfp,daSilveira:2022pte,Xu:2024fun,Shi:2024laj}. In this work, we adopt the so-called weight-RL approximation, which arithmetically averages the strength of interaction for different flavors as follows\,\cite{Qin:2019oar,Xu:2024fun,Shi:2024laj}:
\begin{equation}
 \tilde{\mathcal{D}}^{f\bar{g}}_{\text{eff}} = \eta \mathcal{D}^f_{\text{eff}} + (1-\eta)\mathcal{D}^{\bar{g}}_{\text{eff}}\,.
 \label{eq.wRL}
\end{equation}
Here $\eta$ acts as a weighting factor, controlling the relative strength between the flavor-separated kernels. In the case of flavor-symmetric mesons, Eq.\,\eqref{eq.wRL} reduces to the Eq.\,\eqref{eq:kernel.RL}, hence maintaining the masslessness of the pion in the chiral limit and its Goldstone boson nature. For flavor-asymmetric systems, the effect of weight factor has been discussed in Ref.\,\cite{Qin:2019oar}, in which case it arises naturally from considering certain limits of a more elaborate truncation. In this case, $\eta$ is directly determined from the mass of the pseudo-scalar. This enables us to preserve the simplicities of the RL truncation but better capture the properties of the heavy-light mesons\,\cite{Xu:2024fun,Shi:2024laj}. After the proper determination of the weighting factor $\eta$, the leptonic decay constant of the $D,\ B$ mesons, as well as the mass and leptonic decay constant of the vector counterparts $D^*,\ B^*$, emerge as genuine predictions. As can be seen from Table~\ref{tab:meson}, the resulting values are notably on par with experiment and other phenomenological determinations. Furthermore, an assessment through Eq.\,\eqref{eq.gmor}, indicates a degree of AV-WTI preservation at the level of $\sim 97\%$. \par 
It is worth noting that, under a strict heavy-light kernel, once the flavor-dependence of interaction is determined in the gap equation, fitting for heavy-light mesons is not required. Therefore, the introduction of the weighting factor in this kernel ansatz should be considered as a cost of calculating heavy-light mesons in the RL-like framework, to remedy the effects of flavor asymmetry in BSEs. The theoretical exploration for flavor dependence within strict beyond-RL kernels is still ongoing and remains highly challenging \cite{Ivanov:1998ms,Rojas:2014aka,Gomez-Rocha:2016cji,daSilveira:2022pte,Chen:2019otg,Qin:2020jig}.  \par 
\section{Numerical results and discussion}
\label{sec:3}
With all the above in hand, the strong coupling can be directly calculated based on Eq.\,\eqref{eq.sd}. Due to Poincaré invariance, $g_{H^*H}$ is in principle independent of the choice of $\alpha$ defined in Eq.\,\eqref{eq.krkskp}. However, as previously noted, in numerical computations, the value of $\alpha$ must be strictly constrained by Eq.\,\eqref{eq.alpha} due to the quark propagator's domain of computability, which is influenced by its singularities in the complex plane. Consequently, by varying $\alpha$, we can assess the associated error as an indicator of the preservation of Poincaré invariance in our numerical calculation. \par 

\begin{figure}
\centering 
\includegraphics[width=.35\textwidth]{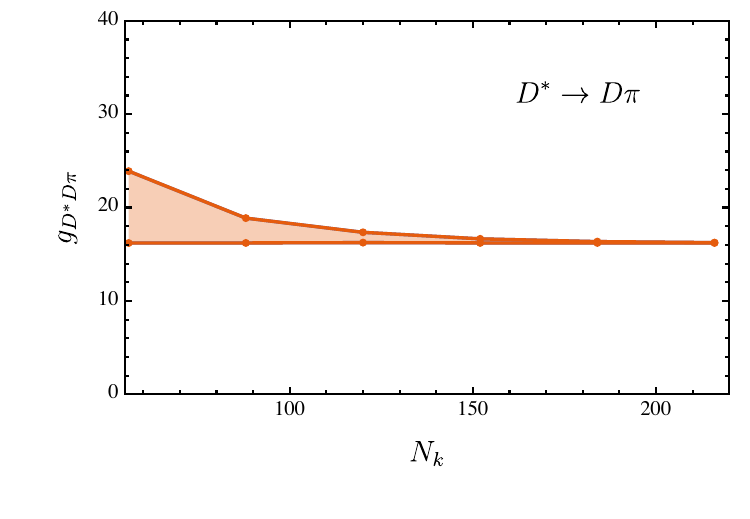}
\includegraphics[width=.35\textwidth]{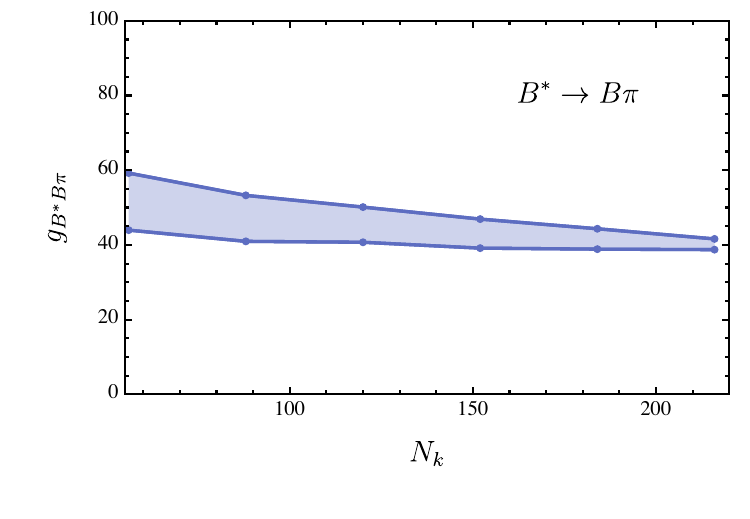}
\caption{\label{fig:sd_alpha} This figure illustrates that the uncertainty in \( g_{H^*H\pi} \), which arises from the variation of  \(\alpha\) within its allowed range (see Eq.\,(\ref{eq.alpha})), progressively decreases as the accuracy of numerical integration improves. Here $N_k$ is the number of radial points of Gaussian numerical integration.}  
\end{figure}
\begin{table*}[!t]
\caption{\label{tab:basis}The contributions of different basis elements (see, Eq.\,\eqref{eq:basis}) to the computed coupling constants. An associated pictorial representation is shown in Fig.\,\ref{fig:basis}.}
\begin{ruledtabular}
\begin{tabular}{llll|llll}
   \multicolumn{4}{c|}{$g_{D^*D\pi}$} & \multicolumn{4}{c}{$g_{B^*B\pi}$} \\[0.5em]
\hline 
 $\tau^{1}_{D^*},\tau^{1}_{D}$ & $29.94_{-0.01}^{+0.05}$ &   $\tau^{1}_{D^*},\tau^{3}_{D}$  & $-0.51_{-0.40}^{+0.01}$ &$\tau^{1}_{B^*},\tau^{1}_{B}$& $90.21_{-2.99}^{+3.33}$ &$\tau^{1}_{B^*},\tau^{3}_{B}$& $-0.50_{-0.19}^{+0.03}$ \\[0.5em]
$\tau^{1}_{D^*},\tau^{2}_{D}$ & $-12.66_{-0.40}^{+0.11}$ &   $\tau^{1}_{D^*},\tau^{4}_{D}$  & $-0.15_{-<0.01}^{+<0.01}$ &$\tau^{1}_{B^*},\tau^{2}_{B}$& $-45.30_{-1.72}^{+1.70}$ &$\tau^{1}_{B^*},\tau^{4}_{B}$& $-0.84_{-0.02}^{+0.02}$ \\[0.5em]
   \hline 
    $\tau^{1}_{D^*}$ & $16.63_{-0.01}^{+0.03}$ &   $\tau^{5}_{D^*}$  & $-2.55_{-<0.01}^{+<0.01}$ &$\tau^{1}_{B^*}$& $43.57_{-1.47}^{+1.63}$ &$\tau^{5}_{B^*}$& $-10.82_{-0.40}^{+0.37}$ \\[0.5em]
 $\tau^{2}_{D^*}$ & $0.19_{-<0.01}^{+<0.01}$ &   $\tau^{6}_{D^*}$  & $-0.13_{-<0.01}^{+<0.01}$ &$\tau^{2}_{B^*}$& $0.41_{-0.02}^{+0.02}$ &$\tau^{6}_{B^*}$& $-0.63_{-0.03}^{+0.03}$ \\[0.5em]
  $\tau^{3}_{D^*}$ & $0.79_{-<0.01}^{+<0.01}$ &   $\tau^{7}_{D^*}$  & $0.25_{-<0.01}^{+<0.01}$ &$\tau^{3}_{B^*}$& $4.79_{-0.17}^{+0.18}$ &$\tau^{7}_{B^*}$& $0.10_{-<0.01}^{+<0.01}$ \\[0.5em]
 $\tau^{4}_{D^*}$ & $1.56_{-<0.01}^{+<0.01}$ &   $\tau^{8}_{D^*}$  & $-0.53_{-<0.01}^{+0.01}$ &$\tau^{4}_{B^*}$& $4.52
   _{-0.17}^{+0.18}$&$\tau^{8}_{B^*}$& $-1.87_{-0.06}^{+0.07}$ \\[0.5em]
   \hline 
 \multicolumn{4}{c|}{$16.22_{-0.01}^{+0.03}$}  &  \multicolumn{4}{c}{$40.09_{-1.37}^{+1.51}$} 
\end{tabular}
\end{ruledtabular}
\end{table*}
In Fig.\,\ref{fig:sd_alpha}, we present the numerical error resulting from variations in the momentum partitioning parameter of vector meson, i.e, $\alpha$. As the number of Gaussian quadrature points in the numerical integration increases, the obtained results become more precise. However, when compared to the rest frame, the memory requirements for solving the BSEs in a moving frame typically increase by two orders of magnitude, reaching the terabyte ($\sim$ TB). While some relevant optimizations can be applied (see appendix \ref{sec.app}), the number of Gaussian integration points is still limited. So, after balancing our computational resources and precision, we report the final result as
\begin{align}
  g_{D^*D\pi} = 16.22_{-0.01}^{+0.03},\ g_{B^*B\pi} = 40.09_{-1.37}^{+1.51}\,. 
\end{align}\par
\begin{figure}
\centering 
\includegraphics[width=.35\textwidth]{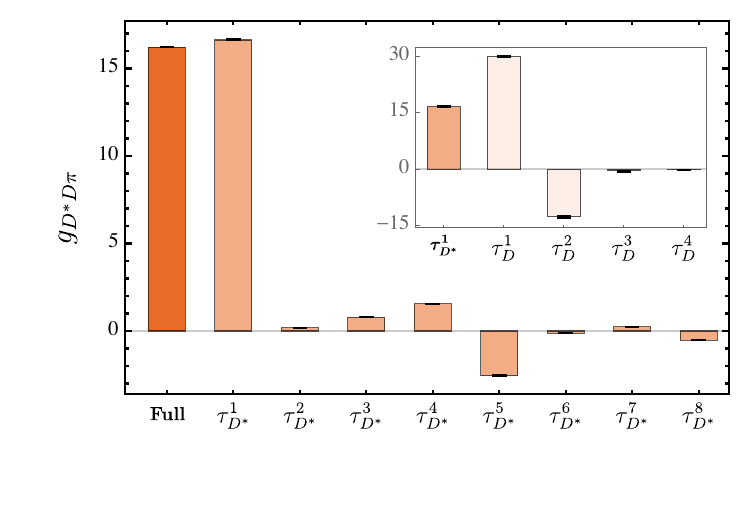}
\includegraphics[width=.35\textwidth]{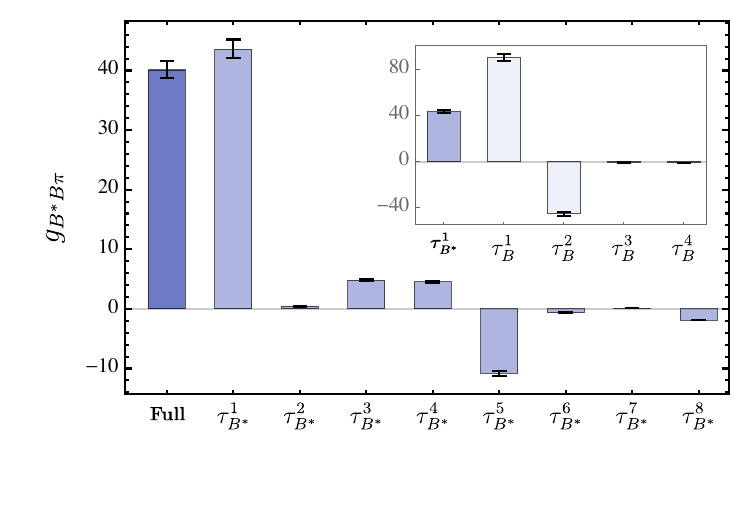}
\caption{\label{fig:basis}The contributions of different basis elements (see, Eq.\,\eqref{eq:basis}) to the computed coupling constants. Further details and the exact values displayed are provided in Table~\ref{tab:basis}.}
\end{figure}

 
It is well-established that the vector mesons BSAs are composed of eight independent Dirac structures (see Eq.\,\eqref{eq:basis}), among which $\tau^{1}_{1^-}$ is usually considered the dominant term. Due to the relatively large number of Dirac covariants, it is worth analyzing the effects of including or neglecting different basis terms. For this purpose, Table~\ref{tab:basis} provides a breakdown of the contributions from different vector BSAs to $g_{H^*H\pi}$, while a complementary visual representation of such is shown in Fig.\,\ref{fig:basis}. \par
To further analyze this dissection, let us define $\{N_{H^*},N_{H}\}$ as the number of Dirac structures in the vector and pseudo-scalar meson BSAs, respectively, considered when evaluating $g_{H^* H \pi}$. Specifically, $N_{H^*,H}=1$ would correspond to retaining only the leading BSA whereas $N_{H^*,H}=8,4$ indicates the consideration of the complete basis. Our numerical results indicate that, when restricting the vector meson BSA to its dominant component, the resulting error in $g_{D^*D\pi}$ is approximately $2.5\%$, while for $g_{B^*B\pi}$, it increases up to $8.7\%$. Both are rather close to the complete results for $\{N_{H^*}=8,N_H={4}\}$, as the contributions from the other terms tend to cancel out each other. This suggests that $\{ N_{H^*}=1,N_{H}=4\}$ provides a fairly good approximation, and the sub-dominant amplitudes of vector meson make only a minor correction, which is consistent with previous explorations concerning the $\rho\pi\pi$ process\,\cite{Tandy:1998ha}. \par 
This being the case, one can focus on retaining only the dominant BSA in the vector meson and thus analyze the contributions of the different pseudo-scalar meson BSAs. In contrast to  the $H^*$ case, the contributions of the first two basis-elements of the $H$ meson BSA are both significant; and, in fact, the $\tau_{0^-}^2$ component (see Eq.\,\eqref{eq:basis}) provides a non-negligible negative correction. Hence, the result for $\{ N_{H^*}=1,N_{H}=1\}$ is approximately twice the complete result. This is not surprising, as the subdominant amplitude $\tau_{0^-}^2$ often plays an important role in accurately characterizing the properties of the pseudo-scalar (for instance, in determining the asymptotic behavior of its electromagnetic form factors),\,\cite{Maris:1998hc}. Although employing a different effective interaction kernel, a similar conclusion for $D^*D\pi$ was reported in Ref.\,\cite{daSilveira:2022pte}, hence supporting our interpretation.\par 
\begin{table*}[!t]
\caption{\label{tab:compare} This table collects our results, experimental values, and some determinations from other approaches. The script ${}^\ddagger$ denotes a calculation based on the formula provided in that paper. The experimental value of $g_{D^*D\pi}$ is derived from the $D^{+} \rightarrow D^0 \pi^{+}$ data provided by PDG\,\cite{ParticleDataGroup:2022pth,Khodjamirian:2020mlb}. Notably, the error in our results presented here reflects only numerical noise arising from the variation of $\alpha$ within its adjustable range. The DSEs/BSEs framework involves multiple sources of uncertainty, and therefore the actual errors in our work are underestimated, especially for $D^*D\pi$.}
\begin{ruledtabular}
\begin{tabular}{lllll}
  & $g_{D^*D\pi}$ & $\hat{g}_D$ & $g_{B^*B\pi}$ & $\hat{g}_B$ \\[0.5em]
\hline 
 This work & $16.22_{-0.01}^{+0.03}$ & $0.55_{-<0.01}^{+<0.01}$ & $40.09_{-1.37}^{+1.51}$ & $0.50_{-0.02}^{+0.02}$ \\[0.5em]
Experiment \cite{ParticleDataGroup:2022pth} & $16.82 \pm 0.25$ & - & - & - \\[0.5em]
 Lattice(16) \cite{Flynn:2015xna} & - & - & ${45.3 \pm 6.0}^\ddagger$ & $0.56 \pm 0.07$ \\[0.5em]
 Lattice(15) \cite{Bernardoni:2014kla} & - & - & - & $0.492 \pm 0.029$ \\[0.5em]
 Lattice(13) \cite{Can:2012tx} & $16.23 \pm 1.71$ & - & - & - \\[0.5em]
 Lattice(13) \cite{Becirevic:2012pf} & $15.8 \pm 0.7$ & $0.53 \pm 0.03$ & - & - \\[0.5em]
 Lattice(12) \cite{Detmold:2012ge,Detmold:2011bp}  & - & - & - & $0.449\pm 0.051$ \\[0.5em]
 DSE(23) \cite{daSilveira:2022pte} & $17.24_{-2.30}^{+3.06}$ & $0.58_{-0.08}^{+0.10}$ & - & - \\[0.5em]
 DSE(11) \cite{El-Bennich:2010uqs} & $15.8_{-10}^{+2.1}$ & $0.53_{-0.03}^{+0.07}$ & $30.0_{-1.4}^{+3.2}$ & $0.37_{-0.02}^{+0.04}$ \\[0.5em]
 SR(21) \cite{Khodjamirian:2020mlb} & $14.1_{-1.2}^{+1.3}$ & - & $30.0_{-2.4}^{+2.6}$ & $0.30_{-0.02}^{+0.02}$ \\[0.5em]
 SR(06) \cite{Duraes:2004uc}&  $17.5 \pm 1.5$ & - & $44.7 \pm 1.0$ & - \\[0.5em]
 SR(01) \cite{Navarra:2001ju}& $14.0 \pm 1.5$ & - & $42.5 \pm 2.6$ & - \\[0.5em]
  HQET(14) \cite{Cheung:2014cka} & $16.8 \mp 0.2$ & $0.566_{+0.008}^{-0.007}$ & $43.9_{-0.2}^{+0.1}$ & $0.540_{-0.002}^{+0.001}$ \\[0.5em]
\end{tabular}
\end{ruledtabular}
\end{table*}

In Table~\ref{tab:compare}, we present our complete results of $g_{H^*H\pi}$ alongside predictions from other theoretical frameworks. Additionally, the value of $g_{D^*D\pi}$ extracted from experimental data is also listed. The PDG reports the total width of $D^{*\pm}$ to be:  
\begin{align}
  \Gamma_{\text {tot }}\left(D^{* \pm}\right)=83.4 \pm 1.8\ \text{keV}\,,
\end{align}
based on BaBar ($83.3 \pm 1.2 \pm 1.4$ keV) and CLEO ($96 \pm 4 \pm 22$ keV) determinations. According to the precisely measured branching fraction, $\mathcal{B} \mathcal{R} (D^{*+} \rightarrow D^0 {\pi^+})= 67.7 \pm 0.5 \%$, one has
\begin{align}
\Gamma_{\text{PDG}}\left(D^{*+} \rightarrow D^0 \pi^{+}\right) &= 56.5 \pm 1.3 \ \text{keV},
\end{align}
with BaBar ($56.4 \pm 1.0$) keV and CLEO ($65 \pm 15)$ keV. Correspondingly, $g_{D^*D\pi}$ can be extracted via the formula\,\cite{Belyaev:1994zk,Khodjamirian:2020mlb}:
\begin{align}
  \Gamma\left(D^{*+} \rightarrow D^0 \pi^{+}\right)&=\frac{g_{D^* D \pi}^2}{24 \pi M_{D^*}^2}k^3,
\end{align}
where
\begin{align}
  k^2= &(M_{D^{*+}}^2-(M_{D^+}+M_{\pi^+})^2) \nonumber \\
  &(M_{D^{*+}}^2-(M_{D^+}-M_{\pi^+})^2)/4M_{D^{*+}}^2 \\
  = &(39.33 \pm 0.24)^2\ \text{MeV}^2\,,
\end{align}
hence arriving at
\begin{align}
  g^{\text{PDG}}_{D^*D\pi} = 16.82 \pm 0.25\,,
\end{align}
whereas one has BaBar $(16.81 \pm 0.2)$ and CLEO $(18.0 \pm 2.1)$.\par 
The $B^*B\pi$ coupling cannot be measured directly, as the decay $B^* \rightarrow B \pi$ is kinematically forbidden. Nevertheless, $g_{B^*B\pi}$ remains of interest because it is closer to the heavy quark limit than $g_{D^*D\pi}$. As mentioned in the introduction, this makes it more suitable for accurately extracting the static strong coupling $\hat{g}$. Following Refs.\,\cite{Mader:2011zf,daSilveira:2022pte,Cheung:2014cka}, we define 
\begin{align}
\hat{g}_H = \frac{g_{H^* H \pi}}{2 \sqrt{M_H M_{H^*}}} f_\pi\,,
\end{align}
and obtain
\begin{align}
  \hat{g}_{D}=0.55_{-<0.01}^{+<0.01},\quad\hat{g}_{B}=0.50_{-0.02}^{+0.02}\,.
\end{align}
In Table~\ref{tab:compare}, the result reported in Ref.\,\cite{daSilveira:2022pte}, denoted as DSE(23), is comparable to ours; therein, a DSEs/BSEs scheme was employed as well, albeit with a different effective heavy-light kernel. This comparison shows that both are close to the experimental values. This agreement is encouraging as the cross-comparison of different effective heavy-light BSE kernels and resolution procedures demonstrates the stability of the current computational framework.\par 
In the case of $g_{B*B\pi}$ and $\hat{g}_{B}$, the reported values from different models exhibit a larger discrepancy. This could be due to the higher flavor asymmetry, which presents theoretical challenges, as observed in the case of electromagnetic form factors\,\cite{Xu:2024fun}. Our result is slightly lower than the central value of $\hat{g}_{B}=0.56\pm 0.07$ from the latest lattice result\,\cite{Flynn:2015xna}, although it is still within the error bar. However, due to the absence of experimental data, these predictions still need to be further verified by different approaches.\par 
It is important to note that in this work, we estimate the uncertainty only by varying $\alpha$ within its adjustable range. This error reflects the background noise inherent in the numerical integration and serves as an indicator of the preservation of Poincaré invariance in our calculations. We eliminate the fitting/extrapolation errors of the BSAs that appeared in previous similar studies in the evaluation of the triangle diagram; however, the DSEs/BSEs framework involves multiple sources of uncertainty. For instance, due to the constraints imposed by singularities of the quark propagator in the complex plane, we are unable to test the stability of the results over a wider range of \(\alpha\). For the same reason, and in line with Ref.~\cite{daSilveira:2022pte}, we refrain from including the uncertainties associated with variations in the parameters of the gluon model. Moreover, the use of the impulse approximation and the effective interaction kernel can also introduce systematic errors that are not readily quantifiable. As a result, the uncertainties presented here certainly underestimate the true values, particularly in the case of the \(D^*D\pi\) process. Further work is required to achieve a more precise determination of the strong decay coupling constants in the DSEs/BSEs framework.
\section{Summary}
\label{sec:4}

In this work, we calculate the strong decay couplings $D^* D \pi$ and $B^* B \pi$ within a DSEs/BSEs based approach. This relies on the impulse approximation, along with a moving frame determination of the involved amplitudes. Notably, the $B^* B \pi$ result is reported for the first time based on a Poincaré invariant determination of the corresponding meson BSAs. \par 
For the $D^*D\pi$ coupling, we predict $g_{D^* D \pi}=16.22_{-0.01}^{+0.03}$ and an associated static strong coupling $\hat{g}_D=0.55_{-<0.01}^{+<0.01}$, which differs from the coupling constant extracted from the PDG experimental data by about 3.5\%. Since the BSEs are solved directly in the moving frame, errors due to fitting/extrapolation can be eliminated. The error reported in this work  arises from the adjustment of $\alpha$, which can be improved with a more accurate numerical integration. Moreover, our results are consistent with those obtained using a similar framework but with a different effective heavy-light BSE kernel. This cross-check demonstrates the stability of the current computational framework. Besides, it should be noted that the uncertainties here underestimate the full systematic errors inherent in the DSEs/BSEs approach.\par 
 Concerning the $B^*B\pi$ case, we predict $g_{B^* B \pi}=40.09_{-1.37}^{+1.51}$ and $\hat{g}_B = 0.50_{-0.02}^{+0.02}$, which are in good agreement with the most recent lattice results. However, in the absence of experimental data for validation, these predictions should be further corroborated through the application of alternative models or approaches. \par 
Finally, we recall that our calculation exploits the impulse approximation, which allows expressing the quantities of interest in terms of quark propagators meson BSAs. These are obtained under a RL-like framework that employs an effective heavy-light kernel. Therefore, further comparisons based on other possible effective kernels are crucial. Moreover, the results should be refined in the future by developing a more advanced approach that goes beyond the RL kernel/impulse approximation. In any case, the scheme herein presented can be straightforwardly applied to investigate other strong decay processes. More broadly, we anticipate that it will be valuable for investigating the static properties and dynamics of QCD's bound states.
 
\begin{acknowledgments}
We would like to thank Lang-Tian Liu for useful discussions. This work is supported by the Spanish MICINN grant PID2022-140440NB-C22, and the regional Andalusian project P18-FR-5057. The authors acknowledge, too, the use of the computer facilities of C3UPO at the Universidad Pablo de Olavide, de Sevilla.  
\end{acknowledgments}

\appendix
\section{Decoupling of BSEs and lossless kernel matrix compression}
\label{sec.app}
As mentioned in Sec.\,\ref{sec:3}, compared to the rest frame, the memory requirements for solving the BSEs in a moving frame typically increase by two orders of magnitude ($\sim$ TB), making related optimizations crucial. We thus rewrite Eq.\,\eqref{eq:hBSE} as
\begin{equation}
\Gamma(k; P)= \int^{\Lambda} \frac{d^4q}{(2\pi)^4}  \tilde{\mathcal{D}}^{\mu\nu}_{\text{eff}}(k-q)\gamma_\mu S\left(q_{+}\right) \Gamma(q; P) S\left(q_{-}\right)  \gamma_\nu,
\label{eq.adx.BSE1}
\end{equation}
where the RL approximation has been applied, and 
\begin{align}
  \Gamma(k; P) = F_i(k,P)\tau_i(k,P)\,.
\end{align}
For convenience, we use the Einstein summation convention in the appendix. Here $\tau_i(k,P)$ defines the basis elements (see, Eq.\,\eqref{eq:basis}) and $F_i(k,P)$ is the associated scalar functions. Therefore Eq.\,\eqref{eq.adx.BSE1} can be expanded as 
\begin{align}
F_i(k,P)\tau_i(k,P) = &\int^{\Lambda} \frac{d^4q}{(2\pi)^4}  \tilde{\mathcal{D}}^{\mu\nu}_{\text{eff}}(k-q)\gamma_\mu S\left(q_{+}\right) \nonumber \\
&F_{\alpha}(q,P)\tau_{\alpha}(q,P)S\left(q_{-}\right)  \gamma_\nu\,.
\label{eq.adx.BSE2}
\end{align}
Multiplying both sides of Eq.\,\eqref{eq.adx.BSE2} by each of the basis elements, and taking the trace over the Dirac indices, yields the following:
\begin{align}
&\text{Tr}(\tau_j(k,P)\tau_i(k,P))F_i(k,P)\nonumber \\
= &\int^{\Lambda} \frac{d^4q}{(2\pi)^4} \tilde{\mathcal{D}}^{\mu\nu}_{\text{eff}}(k-q)\nonumber \\ 
 &\text{Tr}(\tau_j(k,P) \gamma_\mu S\left(q_{+}\right) \tau_{\alpha}(q,P)S\left(q_{-}\right)\gamma_\nu)F_{\alpha}(q,P)\,.
\label{eq.adx.BSE.decoupled}
\end{align}
To express it more concisely, we have:
\begin{equation}
\mathscr{P}^{(1)}_{ji} ({k,P}) F_i(k,P) = \int^{\Lambda} \frac{d^4q}{(2\pi)^4} \mathscr{M}_{j\alpha} ({k,q,P}) F_{\alpha}(q,P),
\label{eq.adx.BSE.decoupled2}
\end{equation}
where we define two matrices which don't have Dirac structure:
\begin{align}
  \mathscr{P}^{(1)}_{ji} ({k,P}) \equiv\ &\text{Tr}(\tau_j(k,P)\tau_i(k,P)),\\
   \mathscr{M}^{(0)}_{j\alpha} ({k,q,P}) \equiv\ &\text{Tr}(\tilde{\mathcal{D}}^{\mu\nu}_{\text{eff}}(k-q) \tau_j(k,P) \nonumber \\
   & \gamma_\mu S\left(q_{+}\right) \tau_{\alpha}(q,P)S\left(q_{-}\right)  \gamma_\nu).
\end{align}
Finally, Eq.\,\eqref{eq.adx.BSE1} is decoupled into
\begin{align}
F_i(k,P) = \int^{\Lambda} \frac{d^4q}{(2\pi)^4} \mathscr{M}^{(1)}_{ij} ({k,P}) \mathscr{M}^{(0)}_{j\alpha} ({k,q,P}) F_{\alpha}(q,P)\,,
\label{eq.adx.BSE.decoupled3}
\end{align}
where $\mathscr{M}^{(1)}(k,P)$ is the inverse matrix of $\mathscr{P}^{(1)}(k,P)$. After discretization using numerical integration, Eq.\,\eqref{eq.adx.BSE.decoupled3} turns into an eigenvalue equation, and the scalar function $F(k, P)$ can be directly obtained. However, the matrix $\mathscr{M}^{(0)}(k, q, P)$ typically does not have non-zero matrix elements. Therefore, it requires a substantial amount of memory resources, particularly in the numerical computations of moving frame BSEs.\par 
Note that when the basis is complete, the BS wave function $\chi(q,P) \equiv S\left(q_{+}\right) \Gamma(q ; P) S\left(q_{-}\right)$  can be expanded in the same basis
\begin{align}
  \chi(q,P) = g_{\alpha}(q,P)\tau_{\alpha}(q,P)\,,
\end{align}
with corresponding $g_{\alpha}(q,P)$ scalar functions. These can be written as
\begin{align}
  g_{\sigma}(q,P) &=  \delta_{\sigma\alpha} g_\alpha(q,P) \\
  =\ &\mathscr{M}^{(1)}_{\sigma \beta}(q,P)  \mathscr{P}^{(1)}_{\beta\alpha}(q,P) g_\alpha(q,P) \\
  =\ &\mathscr{M}^{(1)}_{\sigma \beta}(q,P)\text{Tr} (\tau_{\beta}(q,P) g_{\alpha}(q,P)\tau_{\alpha}(q,P)) \\
  =\ &\mathscr{M}^{(1)}_{\sigma \beta}(q,P)\text{Tr} (\tau_{\beta}(q,P) \chi(q,P))\,, 
\end{align}
such that
\begin{align}
  \chi(q,P) &= g_{\sigma}(q,P)\tau_{\sigma}(q,P)\\
  =\ &\mathscr{M}^{(1)}_{\sigma \beta}(q,P) \text{Tr} (\tau_{\beta}(q,P) \chi(q,P)) \tau_{\sigma}(q,P)\\
  =\ &\mathscr{M}^{(1)}_{\sigma \beta}(q,P) \text{Tr} (\tau_{\beta}(q,P) S\left(q_{+}\right) \Gamma(q ; P)  \nonumber \\
&S\left(q_{-}\right)) \tau_{\sigma}(q,P)\\
  =\ &\mathscr{M}^{(1)}_{\sigma \beta}(q,P)\text{Tr} (\tau_{\beta}(q,P) S\left(q_{+}\right)  \nonumber \\
& F_{\alpha}(q,P)\tau_\alpha(q,P)S\left(q_{-}\right)) \tau_{\sigma}(q,P)\\
  =\ &\mathscr{M}^{(1)}_{\sigma \beta}(q,P) \text{Tr} (\tau_{\beta}(q,P) S\left(q_{+}\right)\tau_\alpha(q,P)S\left(q_{-}\right))
  \nonumber \\
& F_{\alpha}(q,P) \tau_{\sigma}(q,P)\\
  =\ &\mathscr{M}^{(1)}_{\sigma \beta}(q,P)  \mathscr{M}^{(2)}_{\beta\alpha}(q,P) F_{\alpha}(q,P) \tau_{\sigma}(q,P),
\end{align}
where we have defined
\begin{align}
  \mathscr{M}^{(2)}_{\beta\alpha}(q,P) \equiv \text{Tr} (\tau_{\beta}(q,P) S\left(q_{+}\right)\tau_\alpha(q,P)S\left(q_{-}\right))\,.
\end{align}
Therefore, Eq.\,\eqref{eq.adx.BSE1} can be cast as
\begin{align}
\Gamma(k ; P)
= &\int^{\Lambda} \frac{d^4q}{(2\pi)^4} \tilde{\mathcal{D}}^{\mu\nu}_{\text{eff}}(k-q)\gamma_\mu \chi(q,P)\gamma_\nu\\
= &\int^{\Lambda} \frac{d^4q}{(2\pi)^4} \tilde{\mathcal{D}}^{\mu\nu}_{\text{eff}}(k-q)\gamma_\mu \mathscr{M}^{(1)}_{\sigma\beta}(q,P)  \mathscr{M}^{(2)}_{\beta\alpha}(q,P)\nonumber \\
& F_{\alpha}(q,P) \tau_{\sigma}(q,P)\gamma_\nu.
\end{align}
Similar to Eqs.\,\eqref{eq.adx.BSE2}-\eqref{eq.adx.BSE.decoupled3}, the BSAs are expanded and then decoupled, thus producing
\begin{align}
  &\text{Tr}(\tau_{j}(k,P)F_{i}(k,P)\tau_{i}(k,P))\nonumber \\
  = &\int^{\Lambda} \frac{d^4q}{(2\pi)^4} \tilde{\mathcal{D}}^{\mu\nu}_{\text{eff}}(k-q)\text{Tr}(\tau_{j}(k,P) \gamma_\mu \tau_{\sigma}(q,P)\gamma_\nu) \nonumber \\
  & \mathscr{M}^{(1)}_{\sigma\beta}(q,P)  \mathscr{M}^{(2)}_{\beta\alpha}(q,P) F_{\alpha}(q,P) 
  \label{eq.adx.BSE.decoupled.wave}\,.
\end{align}
Subsequently, Eq.\,\eqref{eq.adx.BSE.decoupled.wave} can be written more concisely as
\begin{align}
  \mathscr{P}_{ji}(k,P)F_{i}(k,P)=\ &\int^{\Lambda} \frac{d^4q}{(2\pi)^4} \mathscr{M}_{j \sigma}^{(3)}(k,q,P)\mathscr{M}^{(1)}_{\sigma\beta}(q,P)\nonumber \\
  & \mathscr{M}^{(2)}_{\beta\alpha}(q,P) F_{\alpha}(q,P)\,, 
\end{align}
where
\begin{align}
  \mathscr{M}_{j\sigma}^{(3)}(k,q,P) \equiv \text{Tr}(\tilde{\mathcal{D}}^{\mu\nu}_{\text{eff}}(k-q) \tau_{j}(k,P) \gamma_\mu \tau_{\sigma}(q,P)\gamma_\nu)\,. 
\end{align}
Finally, Eq.\,\eqref{eq.adx.BSE1} is decoupled into
\begin{align}
F_{i}(k,P)= &\int^{\Lambda} \frac{d^4q}{(2\pi)^4}  \mathscr{M}^{(1)}_{ij}(k,P)\mathscr{M}_{j\sigma}^{(3)}(k,q,P)\mathscr{M}^{(1)}_{\sigma \beta}(q,P) \nonumber \\
&\mathscr{M}^{(2)}_{\beta \alpha}(q,P) F_{\alpha}(q,P). 
  \label{eq.adx.BSE.decoupled.wave.end}
\end{align}
Compared to Eq.\,\eqref{eq.adx.BSE.decoupled3}, the decoupling method used in Eq.\,\eqref{eq.adx.BSE.decoupled.wave.end} results in a substantial number of zero-valued matrix elements in the $\mathscr{M}^{(3)}(k, q, P)$, which allows for the storage of only the non-zero elements, thereby saving memory. In this work, memory usage for $4 \times 4$ pseudo-scalar case can be cut down to $6/16\sim 38\%$, while for $8 \times 8$ vector case, it is reduced to $20/64\sim 31\%$. 

\bibliography{ref.bib}

\end{document}